\documentclass[a4paper,english,10pt,superscriptaddress,reprint,amsmath,amssymb,aps,twocolumn,floatfix,notitlepage,showkeys,prd]{revtex4-2}
\usepackage{graphicx}% Include figure files
\usepackage{bm}% bold math
\usepackage{dcolumn}% Align table columns on decimal point
\usepackage[colorlinks=true,citecolor=darkred,urlcolor=darkred, pdfborder={0 0 0}]{hyperref}% add hypertext capabilities
\usepackage{setspace}
\usepackage{caption}
\usepackage{subcaption} % To get subfigures
\usepackage{slashed}
\usepackage[normalem]{ulem}
\usepackage{float}
\usepackage[usenames,dvipsnames]{xcolor}  %%%%%%% usenames,dvipsnames required to get BrickRed
\usepackage{microtype}
\usepackage{lipsum}
\usepackage{amsmath}
\usepackage{amssymb}
\usepackage{mathrsfs}
\usepackage{placeins}
\usepackage{lettrine}   % For Starting a paragraph with a big letter, https://ctan.mines-albi.fr/macros/latex/contrib/lettrine/doc/lettrine.pdf
\input Eileen.fd        % For use of Eileen Caps Reguler from The LaTeX Font Catalogue, https://tug.org/FontCatalogue/eileencapsregular/
\newcommand*\initfamily{\usefont{U}{Eileen}{xl}{n}}

\definecolor{darkred}{rgb}{0.6,0,0}
\definecolor{linkcolor}{rgb}{0,0,0.5}
\usepackage[T1]{fontenc} % if needed
\usepackage[compat=1.1.0]{tikz-feynhand}
\usepackage{tikz-feynman}
\tikzfeynmanset{compat=1.1.0}
\usepackage{feynmp}
\usepackage{tikzsymbols}
\usepackage{array}% For table's line thickness change
\usepackage{pifont} % For special character using commend \ding{}

\definecolor{linkcolor}{rgb}{0,0,0.5}
\usepackage{orcidlink}
\begin{document}
\preprint{2312.xxxx}
\title{\boldmath \color{BrickRed}Extended Scotogenic Model of Neutrino Mass and Proton Decay}
\author{Takaaki Nomura \orcidlink{0000-0002-0864-8333}}
\email{nomura@scu.edu.cn}
\affiliation{College of Physics, Sichuan University, Chengdu 610065, China}
\author{Oleg Popov \orcidlink{0000-0002-0249-8493}}
%\footnotetext{Corresponding author.}
\email{opopo001@ucr.edu (corresponding author)}
\affiliation{Faculty of Physics and Mathematics and Faculty of Biology, Shenzhen MSU-BIT University,\\ 1, International University Park Road, Shenzhen 518172, China}
\date{\today}% It is always \today, today,
\begin{abstract}
The current article presents an extension of the classical scotogenic neutrino mass paradigm, where the three issues in particle physics: dark matter, smallness of neutrino mass, and stability of the proton are interconnected. The scenario encompasses the neutrino mass as well as the proton decay as a consequence of an existence of the dark matter. The study successfully achieves the correlation between the naturally small neutrino masses and naturally long proton lifetime in the present paradigm. Furthermore, all relevant   cosmological, collider, and flavor physics constraints are incorporated in the detailed analysis. The scotogenic fermionic dark matter with the mass in the range from $100$~GeV to $10$~TeV successfully satisfies all relevant constraints. The valid range of $2.51\times 10^{-5} < \lambda < 2\times 10^{-3}$ is obtained for the 2HDM $\lambda$ coupling. We give a brief discussion, as well as, outline some of the future prospects.
\end{abstract}
%Use showkeys class option if keyword display desired
\maketitle % Comment out is "notitlepage" is included in the document definition.
%\tableofcontents
%--------------------------------------------------------------------------------------------------------------------------------------------------------------------------------------------------------
%
\section{Introduction}
\label{sec:intro}
\lettrine[lines=4,findent=-1.3cm]{\normalfont\initfamily \fontsize{17mm}{10mm}\selectfont T \normalfont\initfamily}{ }he origin of neutrino mass is a challenging and long standing problem in particle physics. First thought to be massless, experimental data of observed solar and atmospheric neutrinos from the late 90's changed physicists approach on the problem~\cite{Super-Kamiokande:1998kpq, SNO:2002tuh}. It had been already known, at the time, neutrino oscillation is the one possible phenomena to explain the lack of the solar anti-electron neutrinos~\cite{Cleveland:1998nv,Kamiokande:1996qmi} and anomaly in the atmospheric anti-muon neutrinos~\cite{Kamiokande:1994sgx}. Neutrino oscillations require tiny neutrino masses, indistinguishable by the detectors, and lepton mixing that is described by Pontecorvo-Maki-Nakagawa-Sakata (PMNS) mixing matrix~\cite{Pontecorvo:1957cp,Pontecorvo:1957qd,Maki:1962mu}. As a result oscillation data fixes the neutrino mass splittings and mixing angles of the lepton sector~\cite{ParticleDataGroup:2022pth}. In addition  physicists already had developed mechanisms that would explain the tiny neutrino masses~\cite{Yanagida:1979gs, Minkowski:1977sc, Mohapatra:1979ia,Zee:1980ai}. Starting with tree level~\cite{Yanagida:1979gs, Minkowski:1977sc, Mohapatra:1979ia}, then one loop order~\cite{Zee:1980ai,Ma:1998dn,Ma:2006km}, and later engineering more various models~\cite{Cai:2017jrq} challenging to naturally suppress neutrino mass. The challenge still stands today and we still attempt to come up with naturally suppressed pathways to small neutrino masses~\cite{Ma:1998dn,Ma:2016mwh} and connect, also test, this models via various endless physics, including quark sector, modular symmetries, phase transitions and gravitational waves, topological effects, axions, and other interesting physics.

The other long standing problem in particle physics that remains unresolved since 1970's~\cite{Goldman:1979ij,Donoghue:1979pr,Lazarides:1980nt} is the stability of the proton~\cite{Dorsner:2022twk,Helo:2019yqp,Everett:2021mef,Kang:2024abc}. Since the inception of the grand unified theories (GUTs)~\cite{Georgi:1974sy} and the road to unify all known internal symmetries in nature, the prediction of the proton decay has been a smoking gun signature of any unification scenario. After 70 years of search in the chase of the proton decay signal, the proton decay still remains to be observed. Many experiments~\cite{Super-Kamiokande:2016exg,Irvine-Michigan-Brookhaven:1983iap,ICARUS:2001vle,TITANDWorkingGroup:2001qnw} over the globe build larger detectors and look for more exotic proton decay channels~\cite{Super-Kamiokande:2020tor,Ji-Woong:2023iyc,Super-Kamiokande:2014pqx} in hope to reach the Holy Grail. Several GUT models have since been ruled out, for instance, minimal $SU(5)$~\cite{Georgi:1974yf}. This situation motivated model builders to put forward models with naturally suppressed proton decay, some examples are: $SU(5)$ based models~\cite{Fornal:2017xcj}, models with exotic proton decay channels~\cite{Fonseca:2018ehk,Dorsner:2022twk,Helo:2019yqp,Helo:2018bgb}, radiatively generated proton decays~\cite{Dorsner:2022twk,Helo:2019yqp,Everett:2021mef}, and others~\cite{Everett:2021mef}. As it turned out, creating a GUT model with naturally suppressed proton decay is a no man's land territory. One has to find a smart way: including adding more symmetries, irreducable representations, and {\it so on}; in order to make the proton decay width naturally small or even obtain a completely stable proton.

Under the light of everything discussed above, in the present manuscript we try to put forward the minimal scenario where two aforementioned problems in particle physics are united. We will present a scenario where neutrino mass and proton decay width are naturally suppressed and are correlated. Both achievements require a lepton number violation, hence neutrinos are majorana in nature. Furthermore, both the neutrino mass and proton decay width vanish in the limit when lepton number is conserved. This creates a unique opportunity to complement the neutrino data and the proton decay experimental constraints together and test the enhanced scotogenic neutrino scenario. {\it Scoto} from Greek meaning darkness~\cite{Ma:2006km}. As a bonus, the model also predicts a stable dark matter.

The paper is organized as follows: Sec.~\ref{sec:model} describes the model and its content, neutrino mass generation is given in Sec.~\ref{sec:m_nu}, proton decay and relevant details are studied in Sec.~\ref{sec:p_decay}, constraints on the model's parameters are presented in Sec.~\ref{sec:constraints}, results are given in Sec.~\ref{sec:results}, Sec.~\ref{sec:discussion} contains the discussion, and Sec.~\ref{sec:conclusion} concludes the paper.
\section{Model}
\label{sec:model}
The model presented here is minimal that is capable of generating neutrino mass and proton decay from the common origin, which makes the two correlated. In order to build the model that links the proton decay to the neutrino mass one must go beyond the conventional seesaw scenarios and use fields that simultaneously contribute to the neutrino mass generation and proton decay width. Here, we present a model that adds \emph{ad hoc} $\mathbb{Z}_2$ symmetry, similar to the scotogenic model~\cite{Ma:2006km}, which makes our neutrino mass and proton decay radiative and allows for the dark matter candidates to be present. From one point of view, the model can be viewed as enhanced scotogenic model which extends beyond just neutrino mass generation and encompasses the proton decay width generation as well. At the end, we get a model where the existence of the dark matter is linked to the smallness of the neutrino mass as well as to the smallness of the proton decay width (\emph{i.e.} proton's life time), where the last two are correlated.

The enhanced scotogenic model contains the Standard Model (SM) fields as well as the basic scotogenic fields; Three generations of singlet fermion $N$ and $SU(2)_L$ doublet scalar $\eta$ that are odd under the dark $\mathbb{Z}_2$, while SM fields are even. Furthermore, three generations of the vector-like $D$ electroweak singlet quark and $\Tilde{R}_{2D}$, both odd under the dark $\mathbb{Z}_2$, are part of the enhanced scotogenic model and contribute to the radiative generation of the proton decay width. All fields contained in the model and their transformations under the model's symmetries are given in Tab.~\ref{tab:model_fields}.

Some important properties of the enhanced scotogenic model are both neutrino mass and proton decay width generation is at loop order which contributes to their smallness. Both, the neutrino mass and proton decay width, are proportional to the source of the lepton number violation ($m_N$) and both vanish when the last one approaches zero: no lepton number violation leads to massless neutrinos and stable proton. The details of this correlation are given in the later section of the manuscript.
\begin{table}[h]
    \centering
    \begin{tabular}{ccccc}
        \hline \hline
        Fields & $SU(3)_c$ & $SU(2)_L$ & $U(1)_Y$ & $\mathbb{Z}_2$ \\ \hline
        $Q$ & $\pmb{3}$ & $\pmb{2}$ & $~\frac{1}{6}$ & $+$ \\
        $\Bar{u}$ & $\pmb{\Bar{3}}$ & $\pmb{1}$ & $-\frac{2}{3}$ & $+$ \\
        $\Bar{d}$ & $\pmb{\Bar{3}}$ & $\pmb{1}$ & $~\frac{1}{3}$ & $+$ \\
        $L$ & $\pmb{1}$ & $\pmb{2}$ & $-\frac{1}{2}$ & $+$ \\
        $\Bar{e}$ & $\pmb{1}$ & $\pmb{1}$ & $1$ & $+$ \\
        $N$ & $\pmb{1}$ & $\pmb{1}$ & $0$ & $-$ \\
        $D,\Bar{D}^\dagger$ & $\pmb{3}$ & $\pmb{1}$ & $-\frac{1}{3}$ & $-$ \\ \hline
        $H$ & $\pmb{1}$ & $\pmb{2}$ & $~\frac{1}{2}$ & $+$ \\
        $\Tilde{R}_{2D}$ & $\pmb{3}$ & $\pmb{2}$ & $~\frac{1}{6}$ & $-$ \\
        $\eta$ & $\pmb{1}$ & $\pmb{2}$ & $-\frac{1}{2}$ & $-$ \\ \hline \hline
    \end{tabular}
    \caption{Model field content.}
    \label{tab:model_fields}
\end{table}
Lagrangian for the model fields content (Tab.~\ref{tab:model_fields}) is as follows
\begin{subequations}
\label{eq:lag}
\begin{align}
    \label{eq:lag_yuk}
    -\mathcal{L}^{\text{Yuk}}_{\text{BSM}} &= Y_1 \Bar{D} L \Tilde{R}_{2D} + Y_2 Q N \Tilde{R}_{2D}^\dagger + Y_3 Q D \Tilde{R}_{2D} \\
    &+ Y_4 L N \eta^\dagger + Y_5 Q \Bar{D} \eta \nonumber \\
    &+ m_N N N + m_D D \Bar{D} + \text{h.c.} \nonumber \\
    \label{eq:lag_pot}
    V &= \mu_H^2 \left(H^\dagger H\right) + \mu_\eta^2 \left(\eta^\dagger \eta\right) + \mu_{R}^2 \left(\Tilde{R}_{2D}^\dagger \Tilde{R}_{2D}\right) \nonumber \\
    &+ \lambda_H \left(H^\dagger H\right)^2 + \lambda_\eta\left(\eta^\dagger \eta\right)^2 + \lambda_{R} \left(\Tilde{R}_{2D}^\dagger \Tilde{R}_{2D}\right)^2 \nonumber \\
    &+ \lambda_{R}^\prime \left(\Tilde{R}_{2D}^\dagger \Tilde{R}_{2D} \Tilde{R}_{2D}^\dagger \Tilde{R}_{2D}\right) + \lambda_{H\eta} \left(H^\dagger H\right)\left(\eta^\dagger \eta\right) \nonumber \\
    &+ \lambda_{H\eta}^\prime \left(H^\dagger \eta\right)\left(\eta^\dagger H\right) + \lambda_{HR} \left(H^\dagger H\right)\left(\Tilde{R}_{2D}^\dagger \Tilde{R}_{2D}\right) \nonumber \\
    &+ \lambda_{HR}^\prime \left(H^\dagger \Tilde{R}_{2D}\right)\left(\Tilde{R}_{2D}^\dagger H\right) \nonumber \\
    &+ \lambda_{\eta R} \left(\eta^\dagger \eta\right) \left(\Tilde{R}_{2D}^\dagger \Tilde{R}_{2D}\right) + \lambda_{\eta R}^\prime \left(\eta^\dagger \Tilde{R}_{2D}\right) \left(\Tilde{R}_{2D}^\dagger \eta\right) \nonumber \\
    &+\left( \lambda H H \eta \eta + \lambda_{3R} \Tilde{R}_{2D} \Tilde{R}_{2D} \Tilde{R}_{2D} \eta + \text{h.c.}\right)
\end{align}
\end{subequations}

The inert scalar doublet $\eta$ is written by
\begin{equation}
\eta = \begin{pmatrix} \frac{1}{\sqrt2} (\eta_R + \imath \eta_I) \\ \eta^- \end{pmatrix},
\end{equation}
where $\eta_R^-$ has electric charge $-1$ and the components are mass eigenstates since it does not mix with the SM Higgs.

Here we consider masses of scalar fields from inert doublet $\eta$. After electroweak symmetry breaking, we obtain squared masses of inert scalar fields as follows.
\begin{align}
m_{\eta_{R[I]}}^2 &= \mu^2_\eta + \frac{v^2}{2}(\lambda_{H \eta} + \lambda'_{H \eta} +[-] 2 \lambda), \nonumber \\
m_{\eta^\pm}^2 &= \mu^2_\eta + \frac{v^2 \lambda_{H \eta}}{2},
\end{align}
We thus find mass difference between $\eta_R$ and $\eta_I$ is given by $m_{\eta_R}^2 - m_{\eta_I}^2 = 2 \lambda v^2$. 

Potential minimization condition and SM doublet Higgs mass are given as in the SM
\begin{subequations}
    \label{eq:h_mass_V_min}
    \begin{align}
        \label{eq:V_min}
        0 &= \mu_H^2 + \lambda_H v^2, \\
        \label{eq:h_mass}
        m_{h_R}^2 &= \mu_H^2 + 3 \lambda_H v^2, \\
        m_{h_I}^2 &= m_{h^\pm}^2 = 0, 
    \end{align}
\end{subequations}
where $\left\langle H \right\rangle = v/\sqrt{2}$. 
Also dark leptoquark mass is given as $m_{\tilde{R}_{2D}}^2 = 2 \mu_R^2 + \left( \lambda_{HR} + \lambda_{HR}^\prime \right) v^2$, assuming
$m_R^2 \gg v^2$, and $\lambda_{HR}^{(\prime)}<1$ is taken.

Regarding BSM quarks, $N$ generates its majorana mass via lepton number violating term $m_N N N$~\eqref{eq:lag_yuk} and the mass of the BSM vector-like singlet dark quark, $D$, is $M_D$~\eqref{eq:lag_yuk}. Model contains no BSM gauge bosons or vector particles, while SM gauge bosons obtains their mass via SM Higgs as in the SM scenario.
\section{Neutrino Mass}
\label{sec:m_nu}
\begin{figure}[h]
    \centering
    \includegraphics[width=0.45\textwidth]{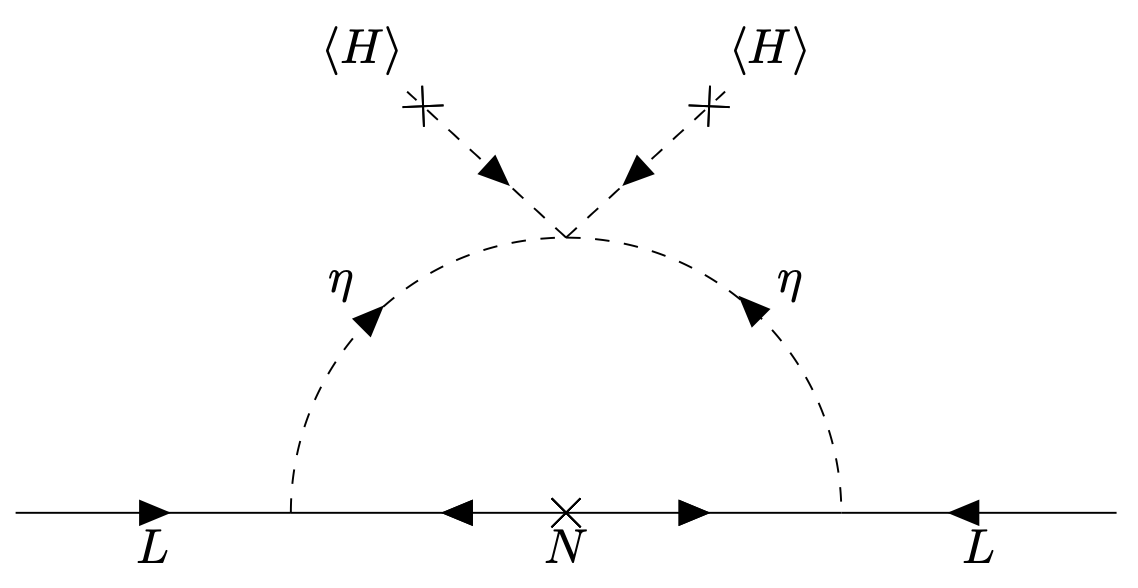}
%    \begin{tikzpicture}
%    \begin{feynman}
%        \vertex (i1);
%        \vertex [right=2cm of i1] (a);
%        \vertex [right=2cm of a] (b);
%        \vertex [right=2cm of b] (c);
%        \vertex [right=2cm of c] (f1);
%        \vertex [above=2cm of b] (t1);
%        \vertex [above left=1.5cm of t1] (tl) {$\left\langle H\right\rangle$};
%        \vertex [above right=1.5cm of t1] (tr) {$\left\langle H\right\rangle$};
%        \diagram* {
%            i1 -- [fermion, edge label'=$L$] (a) -- [anti majorana, insertion=0.5, edge label'=$N$] (c) -- [anti fermion, edge label'=$L$] (f1),
%            a -- [charged scalar, quarter left, edge label=$\eta$] (t1) -- [anti charged scalar, quarter left, edge label=$\eta$] (c),
%            t1 -- [anti charged scalar, insertion=0.9] (tl),
%            t1 -- [anti charged scalar, insertion=0.9] (tr),
%        };
%    \end{feynman}
%    \end{tikzpicture}
    \caption{Scotogenic neutrino mass.}
    \label{fig:_scoto_mnu_2006}
\end{figure}

In the model neutrno mass is radiatively generated by the scotogenic scenario where the relevant one-loop diagram is shown in Fig.~\ref{fig:_scoto_mnu_2006}. 
Calculating the diagram we obtain active neutrino mass matrix such that 
\begin{align}
&(m_\nu)_{ij} = \frac{1}{32 \pi^2} \sum_{k=1}^3 Y_{4}^{ik} m_{N_k} Y_{4}^{jk} \nonumber \\
& \times \left( \frac{m^2_{\eta_R}}{m_{N_k}^2 - m^2_{\eta_R}} \ln \frac{m^2_{\eta_R}}{m^2_{N_k}} -\frac{m^2_{\eta_I}}{m_{N_k}^2 - m^2_{\eta_I}} \ln \frac{m^2_{\eta_I}}{m^2_{N_k}} \right),
\end{align}
where $M_{N_k}$ is a mass eigenvalue of singlet fermion $N$.
If the scale of inert scalar boson mass is much larger than that of $N$, we can approximate $m_\nu$ by
\begin{equation}
(m_\nu)_{ij} \simeq -\frac{1}{16 \pi^2} \frac{\lambda v^2}{M_\eta^2} \sum_{k=1}^3  Y_{4}^{ik} m_{N_k} Y_{4}^{jk},
\label{eq:mnu}
\end{equation}
where $M_\eta^2 = (m^2_{\eta_R} + m^2_{\eta_I})/2$.
In this form we explicitly see that active neutrino masses vanish for $m_{N_k} \to 0$.

\section{Proton Decay}
\label{sec:p_decay}
\begin{figure}[h]
    \centering
    \includegraphics[width=0.45\textwidth]{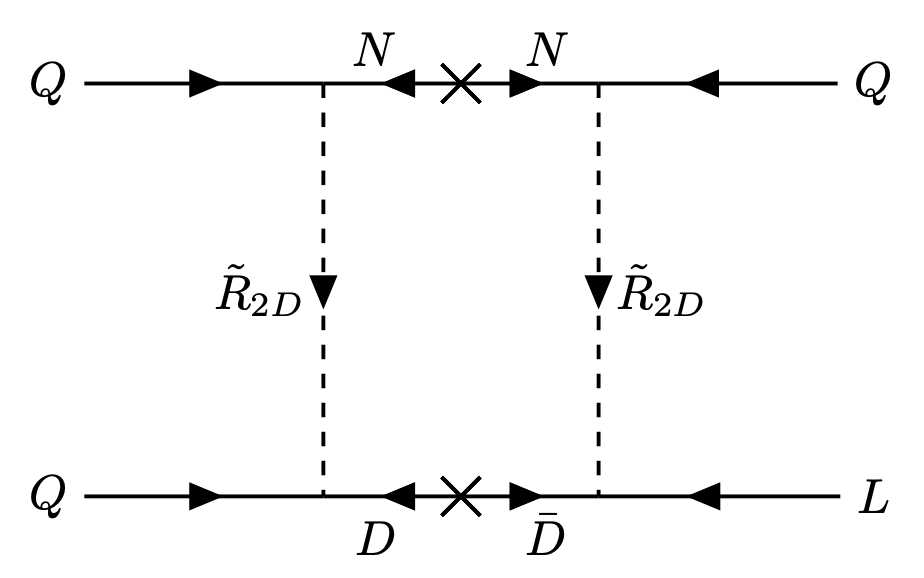}
%    \feynmandiagram [layered layout, horizontal=a to b, large] {
%    i1 [particle=$Q$] -- [fermion] a -- [anti majorana, insertion=0.5, edge label=$N~~~~~~~~N$] b -- [anti fermion] f1 [particle=$Q$],
%    i2 [particle=$Q$] -- [fermion] c -- [anti majorana, insertion=0.5, edge label'=$D~~~~~~~~\Bar{D}$] d -- [anti fermion] f2 [particle=$L$],
%    { [same layer] a -- [charged scalar, edge label'=$\Tilde{R}_{2D}$] c },
%    { [same layer] b -- [charged scalar, edge label=$\Tilde{R}_{2D}$] d},
%};
%Diagram v1, with singlets
% \feynmandiagram [layered layout, horizontal=a to b, large] {
%     i1 [particle=$\Bar{d}$] -- [anti fermion] a -- [majorana, insertion=0.5, edge label=$N~~~~~~~~N$] b -- [fermion] f1 [particle=$\Bar{d}$],
%     i2 [particle=$\Bar{u}$] -- [anti fermion] c -- [majorana, insertion=0.5, edge label'=$\Bar{D}~~~~~~~~D$] d -- [fermion] f2 [particle=$\Bar{\nu}$],
%     { [same layer] a -- [anti charged scalar, edge label'=$\Tilde{\Bar{d}}$] c },
%     { [same layer] b -- [anti charged scalar, edge label=$\Tilde{\Bar{d}}$] d},
% };
    \caption{Feynman diagram contributing to proton decay.}
    \label{fig:p_decay}
\end{figure}

The proton decay is induced at one-loop order through the {\it ad hoc} $\mathbb{Z}_2$ symmetry, similar to the scotogenic scenario. We achieve the radiative proton decay via introducing minimal BSM particle content and as a result get a viable dark matter candidates, which are stabilized by the scotogenic $\mathbb{Z}_2$ symmetry. Since the radiative proton decay resembles the scotogenic structure for the scotogenic neutrino mass scenario, it can be called an enhanced scotogenic mechanism. The radiative nature of the proton decay origin augments the suppression of the width of the long lived proton. Generating the scotogenic proton decay mechanism requires a minimal content of $N$ electrically neutral dark fermion, $\Tilde{R}_{2D}$ dark leptoquark and BSM vectorlike $D$ electroweak (EW) siglet quark. Paricle content is given in Tab.~\ref{tab:model_fields}. Furthermore, due to scotogenic nature of the mechanism, both neutrino mass and proton decay width are proportional to the $N$ fermion's majorana mass. This induces the correlation between the neutrino mass origin and proton decay width generation. Both neutrino mass and proton decay width posses the following characteristics: radiatively suppressed, both are proportional to the lepton number violating term $m_N$ and vanish as $m_N$ approaches zero, both get suppressed as the propagator's mass $m_N$ becomes very large. As a result of the model construction, both neutrino mass and proton decay width are naturally suppressed and are correlated, which creates the opportunity to complement the experimental results from neutrino mass, neutrino oscillations, and proton decay search data to test the enhanced scotogenic model and make further predictions. The radiative proton decay diagram is given in Fig.~\ref{fig:p_decay}. The tree level diagrams are absent due to the implementation of the scotogenic $\mathbb{Z}_2$ symmetry. Feynman diagram in Fig.~\ref{fig:p_decay} contributing to proton decay width and scotogenic neutrino mass feynman diagram in Fig.~\ref{fig:_scoto_mnu_2006} are the leading feynman diagrams generating a correlated scotogenic neutrino mass and proton decay in the framework of an enhanced scotogenic scenario.

In the model at hand proton has two dominant, two body, decay channels, Proton decay via $p \rightarrow e^+ \pi^{0}$ and $p \rightarrow \nu^\dagger \pi^{+}$. The corresponding one loop induced proton decay widths are given as follows
\begin{widetext}
\begin{subequations}
    \label{eq:p_decay}
    \begin{align}
        \Gamma \left(p\rightarrow e^+ \pi^{0},\nu_\alpha^\dagger \pi^{+} \right) &= \frac{m_p}{32\pi} \left(1-\frac{m_\pi^2}{m_p^2}\right)^2 \left| Y_2^{1a} \frac{m_N^{ab}}{m_{\tilde{R}_{2D}}} Y_2^{b1} Y_3^{1c} \frac{m_D^{cd}}{m_{\tilde{R}_{2D}}} Y_1^{d\alpha} \frac{W_0^{l}\left(p\rightarrow e^+ \pi^{0},\nu^\dagger \pi^{+}\right)}{m_{\tilde{R}_{2D}}^2} \right. \\
        &\times\left. \left[ F_0 (x_N, x_D, y) + F_1 (x_N, x_D, y) \frac{q^2}{m_{\tilde{R}_{2D}}^2} \right] \right|^2 , \nonumber
    \end{align}
\end{subequations}
\end{widetext}
where $\alpha = 1$ for $p \rightarrow e^+ \pi^0$, since the constraints on the $p \rightarrow e^+ \pi^0$ are more stringent compared to $p \rightarrow \mu^+ \pi^0$. On the other hand, for neutrinos situation is different. So, $\alpha = 1,2,3$ for $p \rightarrow \nu_\alpha^\dagger \pi^+$, since the neutrino escapes and can be any flavour, while experimental constraint applies to $p \rightarrow \nu^{\left(\dagger\right)} \pi^+$. Loop functions, Kallen $\lambda$, and other definitions are as given below
\begin{widetext}
\begin{subequations}
    \label{eq:fg_functions}
    \begin{align}
    %   F_0 function
        &F_0 (x_N, x_D, y) = \left\{ (1+x_D) (x_N-1) y \ln x_D - (x_D-1)\left[(x_D-x_N) (x_N-1)\ln\left(\frac{x_D}{x_N}\right) - (1+x_N) y \ln x_N \right] \right. \\
        &\left.- (x_D-1)(x_N-1) \lambda^{1/2}(x_D,x_N,y) \left[ \ln(4 x_D x_N) -2 \ln\left(x_D + x_N - y + \lambda^{1/2}(x_D,x_N,y) \right) \right] \right\} \nonumber \\
        &\times \left[ 2 y (x_D-1)(x_N-1) \left((x_D-1)(x_N-1) + y\right) \right]^{-1}, \nonumber \\
    %   F_1 function
        &F_1 (x_N, x_D, y) = \left\{ 4 y \left[ (x_D - x_N)^2 + y \frac{1 - 6 x_D x_N + 2 x_N^2 + 2 x_D^2 + x_D^2 x_N^2}{(x_D-1)(x_N-1)} + y^2 \frac{1 - 4x_D x_N + x_D^2 + x_N^2 + x_D^2 x_N^2}{(x_D-1)^2(x_N-1)^2} \right] \right. \nonumber \\
        &+\left( 2 \left[(x_D-1)^2(x_D-x_N)^2 - 2 y (x_D-1)\left(-x_N + x_D (-2 + x_D + 2 x_N) \right) + y^2 \left( 1 - 4 x_D + x_D^2 \right) \right] \right. \nonumber \\
        &\left.\times \left[ -x_N + y + x_D (1 - x_D + x_N + y) \right] \ln x_D (x_D-1)^{-3} + \left( x_D \leftrightarrow x_N \right) \right) \nonumber \\
        % &+ 2 \left[(x_N-1)^2(x_D-x_N)^2 - 2 y (x_N-1)\left(-x_D + x_N (-2 + x_N + 2 x_D) \right) + y^2 \left( 1 - 4 x_N + x_N^2 \right) \right] \nonumber \\
        % &\times \left[ -x_D + y + x_N (1 - x_N + x_D + y) \right] \ln x_N (x_N-1)^{-3} \nonumber \\
        % &- 2 \left[ x_D^2 + x_N^2 + y^2 - 2 x_N y - 2 x_D y - 2 x_D x_N \right] \lambda^{1/2}(x_D,x_N,y) \nonumber \\
        % &\left.\times \left[ \ln(4 x_D x_N) - 2 \ln \left( x_D + x_N - y + \lambda^{1/2}(x_D,x_N,y) \right) \right] \frac{q^2/m_R^2}{24 y^2 \left[ (x_D-1)(x_N-1) + y \right]^2} \right\}, \nonumber \\
        &\left.- 2 \lambda^{3/2}(x_D,x_N,y) \left[ \ln(4 x_D x_N) - 2 \ln \left( x_D + x_N - y + \lambda^{1/2}(x_D,x_N,y) \right) \right] \right\} \frac{
        %q^2/m_R^2
        1}{24 y^2 \left[ (x_D-1)(x_N-1) + y \right]^2}, \\
        &\hspace{5cm} \lambda\left(a,b,c\right) = a^2 + b^2 + c^2 - 2 a b - 2 a c - 2 b c, \\
        &\hspace{5cm} x_N = \frac{m_N^2}{m_{\tilde{R}_{2D}}^2}, \quad x_D = \frac{m_D^2}{m_{\tilde{R}_{2D}}^2}, \quad y = \frac{m_p^2}{m_{\tilde{R}_{2D}}^2}.
    \end{align}
\end{subequations}
\end{widetext}

Everything defined in dimensionless units for convenience of the reader. Proton decay matrix elements and exchanged momentum between the quarks in the proton are obtained from lattice calculations~\cite{Aoki:2017puj} and are given below

\begin{subequations}
    \label{eq:p_decay_mat_elem}
    \begin{align}
        \label{eq:p_decay_mat_elem_1}
        W_0^e (p \rightarrow e^+ \pi^0) &= \left\langle \pi^0 \right| (ud)_L u_L \left|p\right\rangle \nonumber \\
        &= 0.134(5)(16)~\text{GeV}^2\text{~\cite{Aoki:2017puj}}, \\
        \label{eq:p_decay_mat_elem_2}
        W_0^\nu (p \rightarrow \nu^\dagger \pi^+) &= \left\langle \pi^+ \right| (du)_L d_L \left|p\right\rangle \nonumber \\
        &= 0.189(6)(22)~\text{GeV}^2\text{~\cite{Aoki:2017puj}}, \\
        \label{eq:p_decay_mat_elem_q}
        q^2 &= 0.2~\text{GeV}^2\text{~\cite{Aoki:2017puj}}.
    \end{align}
\end{subequations}

The $W_1^{l}$ for $l=e^+$ or $l=\nu_\alpha^\dagger$ are negligible compared to $W_0^l$ and will be dropped in the further analysis~\cite{Aoki:2017puj}.
\section{Constraints}
\label{sec:constraints}
In this section we summarize relevant observables to constrain the model and corresponding parameters which are limited by them.

%Neutrino mass constraints
The neutrino mass matrix is described by the parameters $\{\lambda, Y_4, m_\eta,  m_N \}$ as Eq.~\eqref{eq:mnu} where $m_\eta = m_{\eta_R} \simeq m_{\eta_I} \simeq m_{\eta^\pm}$. Here we do not reproduce neutrino mixings and require order of $m_\nu$ to be 0.1 eV in order to obtain allowed magnitude of the parameters assuming no hierarchical structure of $Y_4$. 

%Proton lifetime constraints
The most stringent proton lifetime constraint for two body decay was obtained by Super-Kamiokande (SK) collaboration and is given by $\tau(p\rightarrow e^+(\mu^+) \pi^0) > 2.4(1.6)\times 10^{34}$~years~\cite{Super-Kamiokande:2020wjk} and $\tau(p\rightarrow \mu^+ K^0) > 3.6\times 10^{33}$~years~\cite{Super-Kamiokande:2022egr}.
%Super-Kamiokande:2018apg,Super-Kamiokande:2017gev %Old references
Furthermore, the proton decay to any number of final states bound obtained from baryon conservation limit is given by $\tau(p\rightarrow \text{any}) \gtrsim 2\times 10^{30}$~years~\cite{Reines:1974pb}. These constraints apply to the proton decay obtained for our model and that was given in Sec.~\ref{sec:p_decay}. The corresponding results of these constraints are included in the Sec.~\ref{sec:results}.

%LFV contraints
In addition to the constraint from proton decay lifetime, we also consider lepton flavor violations (LFVs) in the scotogenic model. As the most stringent constraint we focus on $\ell \to \ell' \gamma$ process in this work. The branching ratio(BR) of the process is given by~\cite{Toma:2013zsa}
\begin{align}
\label{eq:lfv_br}
\rm{Br}(\ell \to \ell' \gamma) =& \frac{3 (4 \pi)^3}{4 G_F^2} |(A_D)_{\ell \ell'}|^2 {\rm Br}(\ell \to \ell' \nu_{\ell} \bar{\nu}_{\ell'}), \\
(A_D)_{\ell \ell'} = & \sum_{i=1}^3 \frac{(Y_4^*)^{i \ell'} Y_4^{i \ell} }{2 (4 \pi)^2} \frac{1}{m_{\eta^\pm}^2} F_2 \left( \frac{M_{N_i}^2}{m_{\eta^\pm}^2} \right),
\end{align}
where $F_2(x)$ is a function obtained from loop momentum integration:
\begin{equation}
F_2(x) = \frac{1-6x + 3 x^2 + 2 x^3 - 6 x^2 \ln x}{6(1-x)^4}.
\end{equation}
In our analysis, we consider elements of Yukawa coupling $Y_4$ have similar size, and apply the strongest bounds from $\mu \to e \gamma$~\cite{MEG:2016leq} to estimate typical allowed scale of masses $\{m_{N_i}, m_{\eta^\pm} \}$ and the magnitude of $Y_4$.
We thus adapt 
\begin{equation}
{\rm Br}(\mu \to e \gamma) < 4.2 \times 10^{-13}.
\end{equation}
%

%CMS LHC bounds on LQ
We also take into account a constraint on the leptoquark mass $m_{\Tilde{R}_{2D}}$ from collider experiments.
Firstly, it should be noted that the leptoquark in the model has $Z_2$ odd parity and its decay mode is different from those of typical leptoquarks which are searched for at the LHC.
In this case we consider components of $\tilde R_{2D}$ decays into $q N$ mode with $q$ being a SM quark.
Thus, the signal is two jets (heavy quarks) with missing transverse energy that is obtained from pair production of the leptoquarks at the LHC.
Therefore, relevant constraints can be obtained from squark searches~\cite{CMS:2019zmd}.
The analysis of squark searches indicate lower limit of the mass is around 1130~GeV for one type of squark pair production decaying into jet and neutralino (missing energy) when the mass difference between the squark snd neutralino is sufficiently large.
We adopt the constraint for our leptoquark mass.

\begin{figure}
    \centering
    \begin{subfigure}[b]{0.45\textwidth}
    \includegraphics[width=0.45\textwidth]{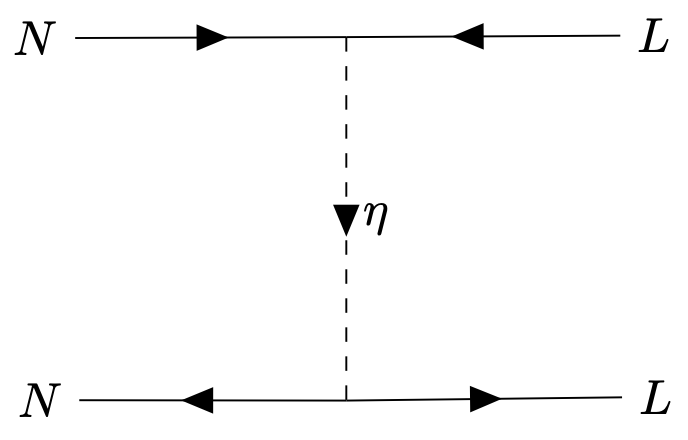}
%    \feynmandiagram [vertical=a to b] {
%    i1 [particle=$N$] -- [fermion] a -- [anti fermion] f1 [particle=$L$],
%    a -- [charged scalar, edge label=$\eta$] b,
%    i2 [particle=$N$] -- [anti fermion] b -- [fermion] f2 [particle=$L$],
%    i1 -- [plain, opacity=0] i2,
%    f1 -- [plain, opacity=0] f2,
%    };
    \caption{DM annihilation to SM leptons via $\eta$ mediated t-channel Feynmann diagram. $\eta$ contribution to DM relic density.}
    \label{fig:dm_L_cont}
    \end{subfigure}
    \\
    \begin{subfigure}[b]{0.45\textwidth}
    \includegraphics[width=0.45\textwidth]{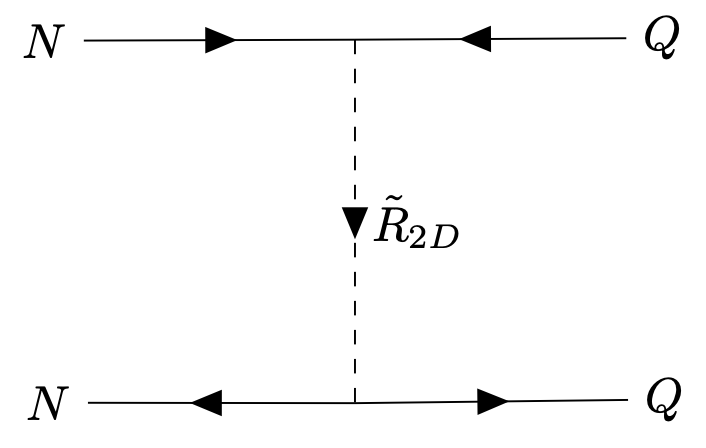}
%    \feynmandiagram [vertical=a to b] {
%    i1 [particle=$N$] -- [fermion] a -- [anti fermion] f1 [particle=$Q$],
%    a -- [charged scalar, edge label=$\Tilde{R}_{2D}$] b,
%    i2 [particle=$N$] -- [anti fermion] b -- [fermion] f2 [particle=$Q$],
%    i1 -- [plain, opacity=0] i2,
%    f1 -- [plain, opacity=0] f2,
%    };
    \caption{DM annihilation to SM quarks via $\Tilde{R}_{2D}$ mediated t-channel Feynmann diagram. $\Tilde{R}_{2D}$ contribution to DM relic density.}
    \label{fig:dm_Q_cont}
    \end{subfigure}
    \caption{Feynmann diagrams contributing to the DM annihilation cross-section. $\eta$ and $\Tilde{R}_{2D}$ contributions.}
    \label{fig:dm_ab_coef}
\end{figure}

%DM constraints, Relic and DD
The presence of DM candidates in the model sets DM relic and Direct Detection (DD) constraints. There are only two possible DM candidates $N$ and $\eta^0$. We consider $N$ as the main DM candidate. The relevant equations for obtaining the DM relic abundance are given by
\begin{subequations}
    \label{eq:dm_const}
    \begin{align}
        \label{eq:dm_relic}
        \Omega_N h^2 &\approx \frac{1.07 \times 10^9 \text{GeV}^{-1}}{M_{\text{pl}}}\frac{x_F}{\sqrt{g_\star}}\frac{1}{\left(a + 3 b /x_F\right)}, \\
        \label{eq:dm_xf}
        x_F &= \ln \left[ c(c+2) \frac{45}{8} \frac{g_N}{2\pi^3} \frac{m_N M_{\text{pl}} (a+6b/x_F)}{\sqrt{g_\star} \sqrt{x_F}} \right], \\
        \label{eq:sv_nonrel_exp}
        \left\langle \sigma v \right\rangle &\approx a + b \left\langle v^2 \right\rangle + \mathcal{O} \left( \left\langle v^4 \right\rangle \right) \approx a + 6 b / x,
    \end{align}
\end{subequations}
where $x=m_N/T$, $F$ index stands for freeze-out temperature or $x_F$ value, $c$ is the order one constant determined from matching the early-time and late-time solutions of eq.~\eqref{eq:dm_xf}. 
$M_{\text{pl}}=\sqrt{\hbar c /G_N} = 1.220 \times 10^{19}$~GeV$/c^2$~\cite{ParticleDataGroup:2022pth} is the Planck mass. 
% $M_{\text{pl}}=\sqrt{\hbar c /G_N} = 2.435 \times 10^{18}$~GeV$/c^2$~\cite{ParticleDataGroup:2022pth} is the reduced Planck mass. 
Eq.~\eqref{eq:sv_nonrel_exp} is the non-relativistic approximation of the DM annihilation cross-section. $a$ and $b$ have the units of GeV$^{-2}$ and are given explicitly for our model in Eqs.~\eqref{eq:dm_ab_coef}. $g_N=2$ is the degrees of freedom (DoF) of the DM particle and $g_\star$ are the relativistic DoF, in thermal equilibrium with SM sector, at the time or temperature of DM freeze-out and $h=H_0 / 100$ km s$^{-1}$ Mpc$^{-1}$, where $H_0=67.4\pm 0.5(73.0\pm 1.0)~\text{km~s}^{-1}~\text{Mpc}^{-1}$ from CMB Planck~\cite{Planck:2018vyg} (SH0ES~\cite{Riess:2021jrx}) is the Hubble parameter of the universe today. Eq.~\eqref{eq:dm_relic} determines the DM relic abundance for a given $a$ and $b$ coefficients, determined by the model (Eq.~\ref{eq:dm_ab_coef}). Most up-to-date results on DM relic abundance are those by Planck Collaboration, $\Omega_N h^2 \approx 0.1200(12)$ (2018)~\cite{Planck:2018vyg}. $a$ and $b$ coefficients are generated by Feynmann diagrams given in Fig.~\ref{fig:dm_ab_coef} and are explicitly given as follows

\begin{subequations}
    \label{eq:dm_ab_coef}
    \begin{align}
        \label{eq:dm_a_coef}
        a = &\frac{9 m_N^2 Y_4^4}{32 \pi \left( m_N^2 + m_\eta^2 \right)^2 } + \frac{3 m_N^2 Y_2^4}{8 \pi \left( m_N^2 + m_{\Tilde{R}_{2D}}^2 \right)^2 }, \\
        \label{eq:dm_b_coef}
        b = &-\frac{3 m_N^2 Y_4^4 \left( 11 m_N^4 + 30 m_N^2 m_\eta^2 - 5 m_\eta^4 \right) }{64 \pi \left( m_N^2 + m_\eta^2 \right)^4 } \nonumber \\
        &- \frac{ m_N^2 Y_2^4 \left( 11 m_N^4 + 30 m_N^2 m_{\Tilde{R}_{2D}}^2 - 5 m_{\Tilde{R}_{2D}}^4 \right) }{16 \pi \left( m_N^2 + m_{\Tilde{R}_{2D}}^2 \right)^4 }.
    \end{align}
\end{subequations}

The results for the DM relic density satisfying the experimental data of~\cite{Planck:2018vyg} are included in the plots of Figs.~\ref{fig:1_mn_meta} and~\ref{fig:2_mn_mR} in Sec.~\ref{sec:results}.

Next, DM DD constraint is considered. DM DD contribution in the model comes from crossing the diagram shown in Fig.~\ref{fig:dm_Q_cont} with the $\Tilde{R}_{2D}$ LQ in the s-channel. The contribution to the DM DD cross-section from Feynmann diagram in Fig.~\ref{fig:dm_Q_cont} is given by~\cite{Belanger:2008sj}
\begin{subequations}
    \label{eq:dm_dd}
    \begin{align}
        \sigma_{\text{nucleon}}^{\text{SI}} &= \frac{\mu_N^2}{\pi}\left[\displaystyle\sum_{q=u,d,s}\left( f_q^p + f_q^n \right)\right]^2 \frac{Y_2^4}{\left( s - m_{\Tilde{R}_{2D}}^2\right)^2}, \\
        s &\simeq m_N^2 \left( 1 + \frac{m_{p,n}}{m_N} \right)^2, \\
        \mu_N &= \frac{m_N m_{p,n}}{m_N + m_{p,n}},
    \end{align}
\end{subequations}
where $p$ and $n$ correspond to proton and neutron, respectively. Meanwhile, $f_{u,d,s}^{p,n}$ matrix elements are (\cite{Belanger:2008sj} and references therein)
\begin{subequations}
    \label{eq:matrix_elements}
    \begin{align}
        &f_d^p = 0.033,\quad f_u^p = 0.023,\quad f_s^p = 0.26, \\
        &f_d^n = 0.042,\quad f_u^n = 0.018,\quad f_s^n = 0.26.
    \end{align}
\end{subequations}
The most stringent up-to-date limit on the DM DD corresponds to the spin-independent DM-nucleon scattering cross-section, excluding cross-sections stronger than $9.8\times 10^{-48}$~cm$^2$ at the DM mass of $36$~GeV/c$^2$ presented by the LUX-ZEPLIN data~\cite{LZ:2022lsv}.

Using the constraints listed in the current section, the bounds on the model parameters are derived and summarized in the following section. 
\section{Results}
\label{sec:results}
\begin{figure}[ht]
    \centering
    \includegraphics[width=0.45\textwidth]{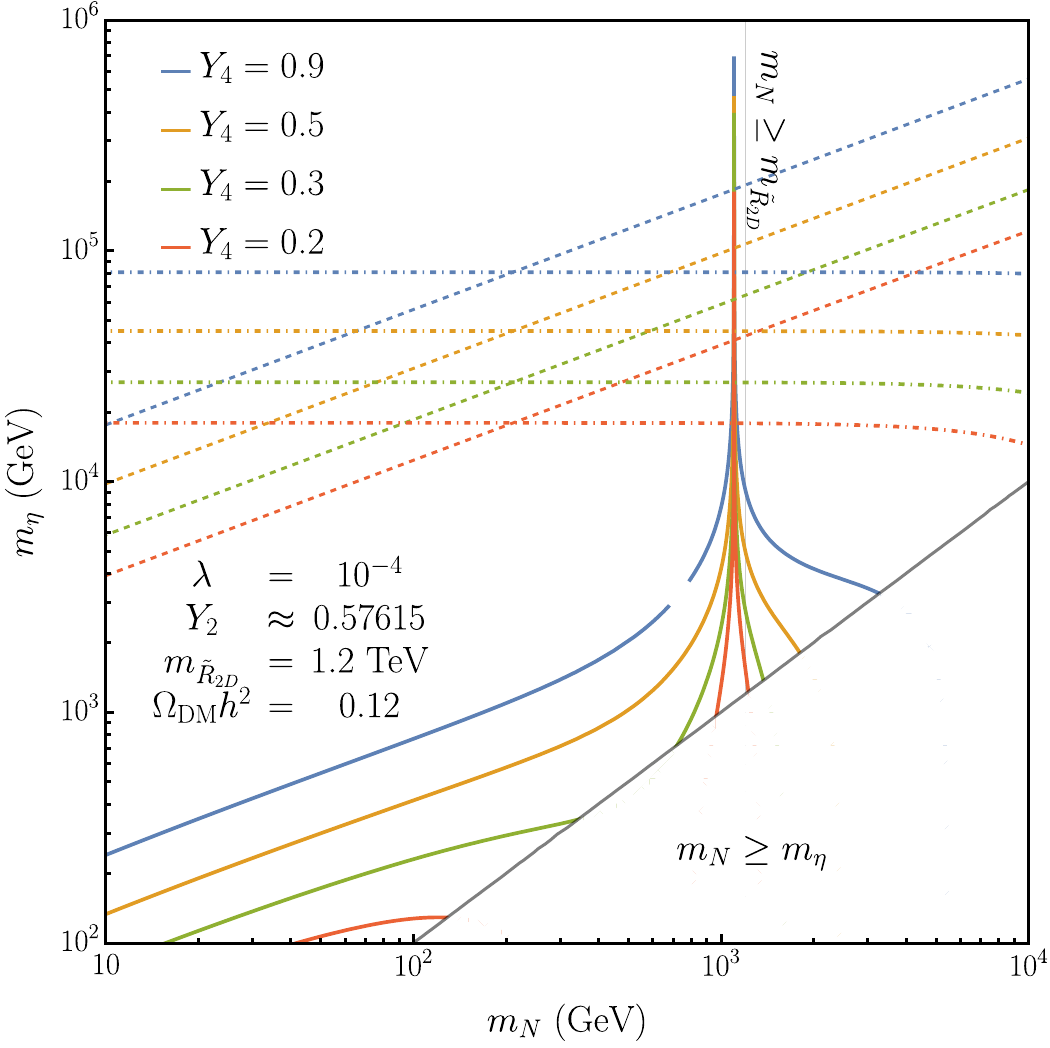}
    \caption{$m_N$ vs $m_\eta$ plot. Solid contours correspond to the correct relic DM density for Yukawa $Y_4$ couplings listed in the plot. Dashed colored contours represent the experimental neutrino mass value for Yukawa $Y_4$ couplings listed in the plot. The dash-dotted contours signify the lower mass bound on the $m_\eta$ obtained from the Br$(\mu\rightarrow e\gamma)<4.1\times 10^{-13}$~\cite{MEG:2016leq} LVF constraints. Since $N$ is DM, $m_N< m_{\eta,\Tilde{R}_{2D}}$ limits are included as well.}
    \label{fig:1_mn_meta}
\end{figure}

The outcomes are summarised in the two following plots. The Fig.~\ref{fig:1_mn_meta} presents a plot of DM mass, $m_N$, vs inert Higgs doublet mass, $m_\eta$, together with incorporated DM relic (Eq.~\eqref{eq:dm_relic}), neutrino mass, and LFV branching ratio(Eq.~\eqref{eq:lfv_br}) constraints. The plot in Fig.~\ref{fig:1_mn_meta} contains the following information: colors of the contours correspond to the values of the Yukawa coefficients, $Y_4$ from Eq.~\eqref{eq:lag_yuk}, listed in the legend, solid contours correspond to the observed DM relic density, dashed diagonal lines represent the valid neutrino mass scale with $\lambda$ value given in the plot, finally dash-dotted flat contours indicate the lower mass bound on the $m_\eta$ set by the LFV constraint. Furthermore, plot includes the mass hierarchy constraints $m_N \geq m_{\eta,\Tilde{R}_{2D}}$. The Yukawa $Y_4$ values were chosen of the order $\mathcal{O}(0.1)$ due to DM relic density constraint. Lower $Y_4$ values $\lesssim\mathcal{O}(10^{-2})$ lead to an overabundant DM relic density. For a given set of Yukawa values (Fig.~\ref{fig:1_mn_meta}), the allowed $\lambda$(Eq.~\eqref{eq:mnu}) range, constrained by the LFV and DM relic, is $\lambda \in (2.51\times 10^{-5}, 2\times 10^{-3})$. If $\lambda < 2.51\times 10^{-5}$ LFV constraint is violated, on the other hand, if $\lambda > 2\times 10^{-3}$ the DM relic cannot be satisfied. DM relic sets a lower bound on the $Y_2\approx 0.576$ Yukawa value.

\begin{figure}[ht]
    \centering
    \includegraphics[width=0.45\textwidth]{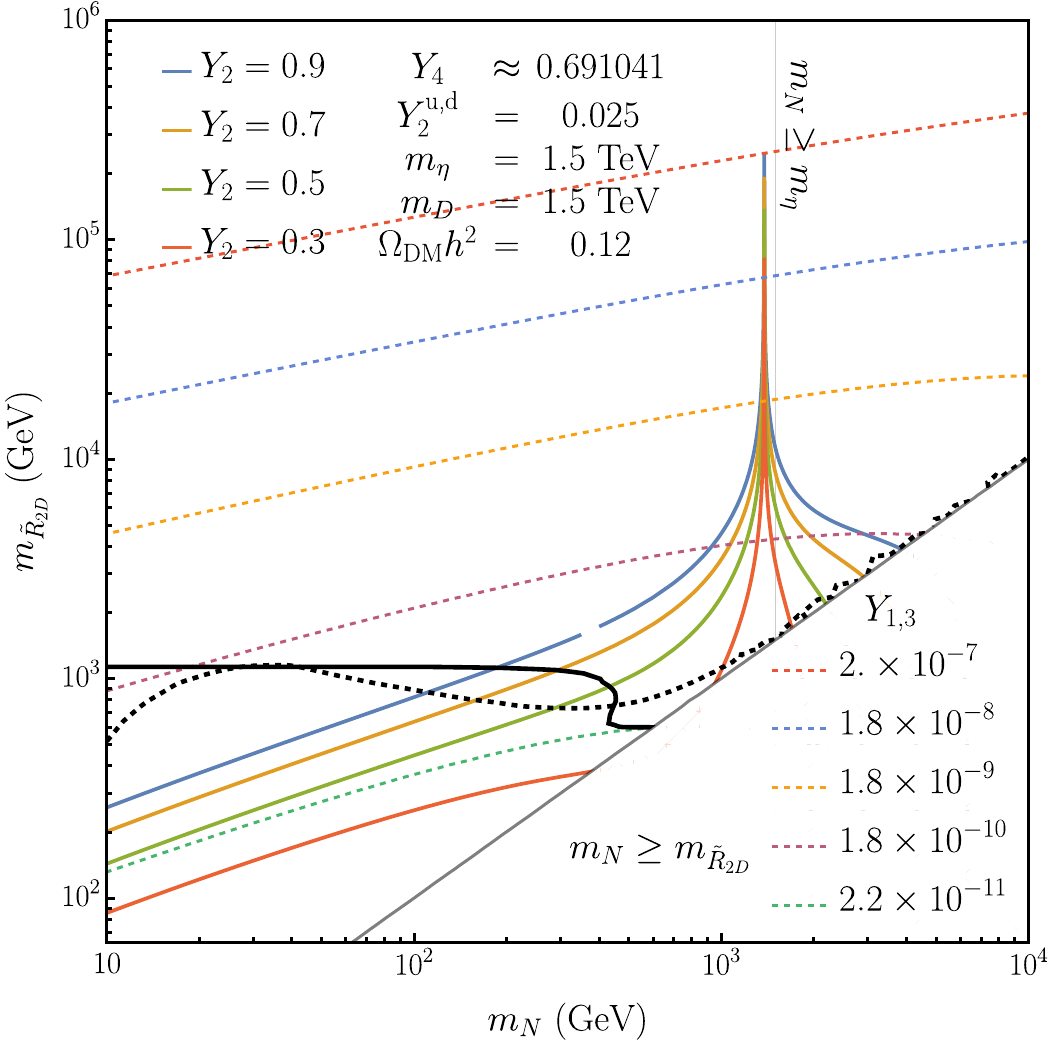}
    \caption{$m_N$ vs $m_{\Tilde{R}_{2D}}$ plot. Solid black contour corresponds to the CMS LHC~\cite{CMS:2019zmd} bound on Leptoquark $\Tilde{R}_2$. Dashed black contour identifies the DM DD constraints obtained from LUX-ZEPLIN (LZ) Experiment~\cite{LZ:2022lsv} 2022 Data with $Y_2^{u,d}=0.025$. Solid contours correspond to the correct relic DM density for Yukawa $Y_2$ couplings listed in the plot. Dashed colored contours represent the \emph{lower bound} on the $m_{\Tilde{R}_{2D}}$ obtained from experimental proton lifetime constraints for $Y_{1,3}$ couplings listed in the plot. Since $N$ is DM, $m_N< m_{\eta,\Tilde{R}_{2D}}$ limits are included as well.}
    \label{fig:2_mn_mR}
\end{figure}

The plot in Fig.~\ref{fig:2_mn_mR} demonstrates the dependence of proton lifetime (Eq.~\eqref{eq:p_decay}), DM relic density (Eq.~\eqref{eq:dm_relic}), DM DD (Eq.~\eqref{eq:dm_dd}), and collider constraints on the DM mass, $m_N$, and LQ mass $m_{\Tilde{R}_{2D}}$. The Fig.~\ref{fig:2_mn_mR}'s plot comprises the following information: solid contours correspond to the observed DM relic density with colors of the solid contours corresponding to the values of the Yukawa coefficients, $Y_2$ from Eq.~\eqref{eq:lag_yuk}, listed in the legend, dashed contours represent the lower $\Tilde{R}_{2D}$ LQ mass bound set by the proton lifetime experimental limits~\cite{Super-Kamiokande:2020wjk,Super-Kamiokande:2022egr,Reines:1974pb} with colors of the dashed contours corresponding to the values of the Yukawa coefficients, $Y_{1,3}$, from Eq.~\eqref{eq:lag_yuk}, black solid line corresponds to the CMS LHC constraint~\cite{CMS:2019zmd} on the $\Tilde{R}_{2D}$ LQ mass, finally black dashed contour indicate the lower mass bound on the $m_{\Tilde{R}_{2D}}$ set by the DM DD constraint (Eq.~\eqref{eq:dm_dd}). Furthermore, plot includes the mass hierarchy constraints $m_N \geq m_{\eta,\Tilde{R}_{2D}}$. One important observation is the upper limit of $Y_{1,3}$ Yukawas set by the proton lifetime and DM relic constraints for $Y_4\approx 0.69$. Furthermore, if $Y_4 < 0.69$ then the upper limit on the $Y_{1,3}$ Yukawas tightens down to $2.2\times 10^{-11}$.

The DM DD requires the Yukawa coupling, $Y_2^{u,d}$, of the DM to the first family SM quakrs to be of the order $\mathcal{O}(0.01)$ in order to satisfy the constraint. On the other hand, the DM abundance forces Yukawas of the order of $\mathcal{O}(0.1)$ for the DM couplings to the SM leptons and second and third family quakrs to obtain the observed DM relic density.

The next generation proton decay search experiments, such as Hyper-Kamiokande~\cite{Hyper-Kamiokande:2018ofw}, Deep Underground Neutrino Experiment (DUNE)~\cite{DUNE:2020fgq}, and \emph{paleo-detectors} proposed in~\cite{Baum:2024sst}, will probe the proton lifetime up to the $10^{34}-10^{35}$~yr values~\cite{Hyper-Kamiokande:2018ofw,DUNE:2020fgq,Baum:2024sst}. Thereby, further restricting the parameter space of the model which will play an important role on the validity of the current model in the forthcoming future.

\begin{figure}[ht]
    \centering
    \includegraphics[width=0.45\textwidth]{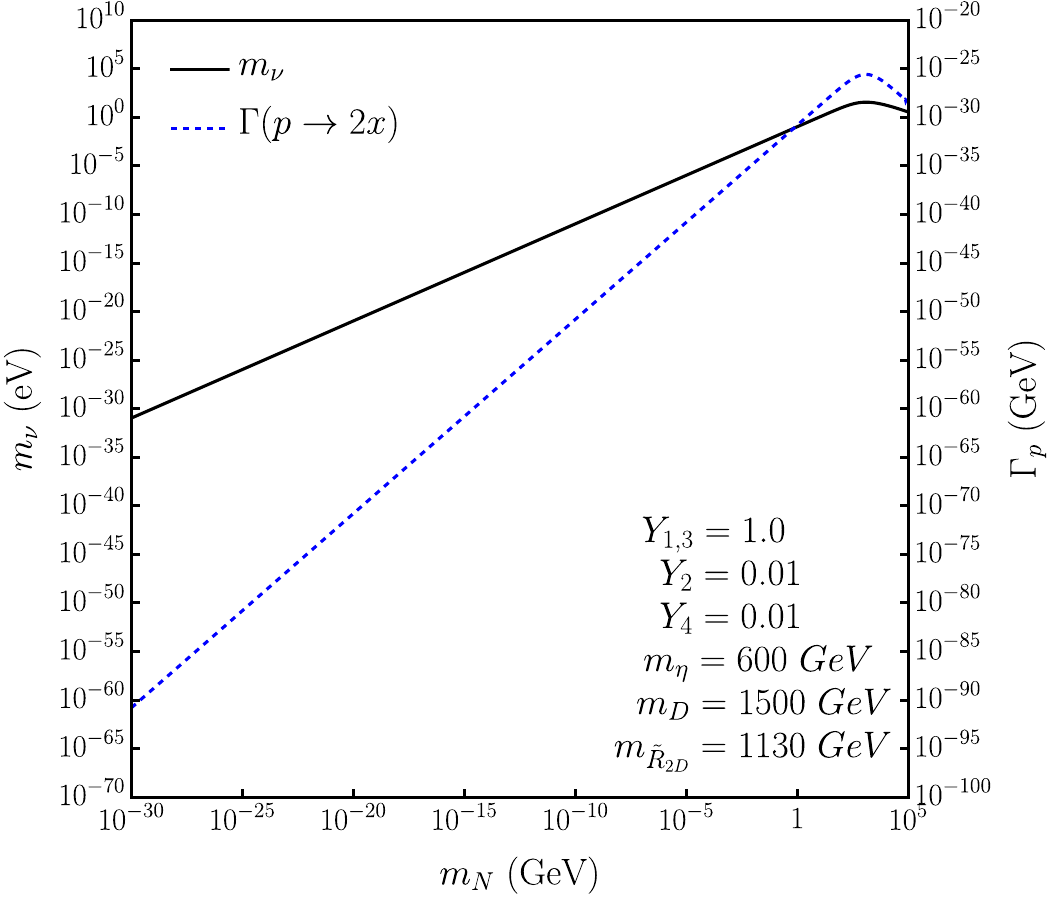}
    \caption{$m_N$ vs neutrino mass and proton decay width correlation plot with other particle masses and Yukawa couplings fixed to the values given in the plot.}
    \label{fig:3_corr_plt}
\end{figure}

One of the main properties of the model is the natural smallness of the neutrino mass and naturally long lifetime of the proton. Furthermore, the model predicts a correlation between the smallness of the two quantities: neutrino mass and proton decay width. This correlation is explicitly plotted in the Fig.~\ref{fig:3_corr_plt} for a given benchmark point of the model parameters. These parameters are included in the Fig.~\ref{fig:3_corr_plt}.

\begin{figure}[ht]
    \centering
    \includegraphics[width=0.45\textwidth]{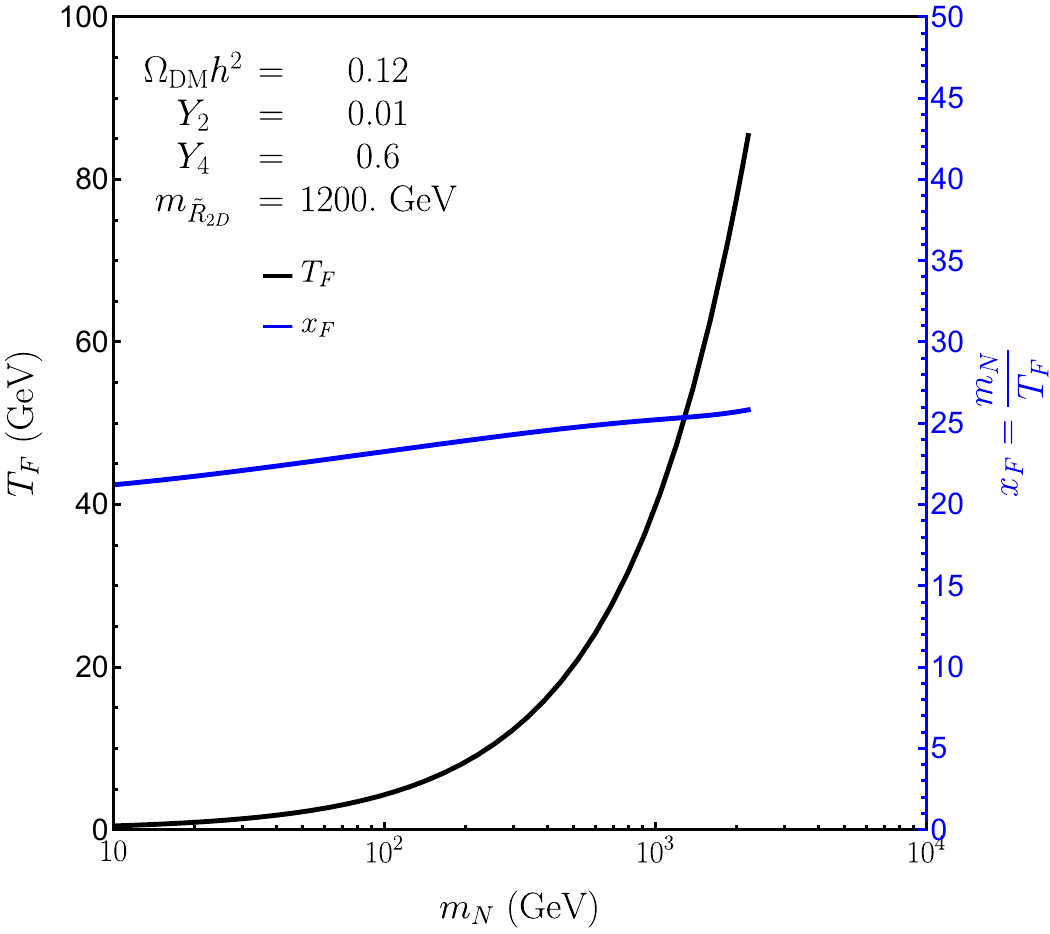}
    \caption{$m_N$ vs freeze-out temperature ($T_F$) and $x_F=\frac{m_N}{T_F}$ plot satisfying the experimental DM relic density constraint with $m_{\Tilde{R}_{2D}}$ LQ mass and Yukawa couplings fixed to the values given in the plot.}
    \label{fig:4_freeze-out_plt}
\end{figure}

The final plot in Fig.~\ref{fig:4_freeze-out_plt} demonstrates the DM freeze-out temperature, $T_F$, and $x_F = m_N/T_F$ parameter for a range of DM masses, $m_N$, and a given benchmark point of the model parameters, which are incorporated in the Fig.~\ref{fig:4_freeze-out_plt} as well.

\section{Discussion}
\label{sec:discussion}
Here we discuss our findings.
Firstly, constraints from neutrino mass and LFV restricts parameters $\{m_N, m_\eta, Y_4, \lambda \}$ as summarized in Fig.~\ref{fig:1_mn_meta}. 
In addition interactions associated with $Y_4$ contribute to DM annihilation $N \bar N \to \ell \bar \ell (\nu \bar \nu)$ via $\eta$ exchange. 
We find it is difficult to explain observed relic density only by the interactions due to LFV constraint requiring small $Y_4$ or heavy $\eta$; 
one can consider large hierarchy of $Y_4$ couplings to avoid LFV constraint but it would not be favored for neutrino sector to fit experimental data causing hierarchy among elements of neutrino mass matrix.
We find that the process $N \bar N \to q \bar q$ via leptoquark $\Tilde{R}_{2D}$ exchange assists in the explanation of the observed relic density. The strength of this process is characterized by the coupling $Y_2$.

The parameters $\{m_N, m_\eta, m_{\tilde{R}_{2D}}, Y_{1,2,3} \}$ are also constrained by proton decay, DM direct detection, collider experiments and relic density as summarized in Fig.~\ref{fig:2_mn_mR}.
For $Y_2$ coupling, hierarchy among components are required to satisfy collider, direct detection and proton decay constraints, explaining relic density at the same time; we need to make $Y_2$ components associated with first generation quarks to be much smaller the other components. Also proton decay constraints require smallness of $Y_{1,3}$ couplings.

In order to achieve the correlation between the smallness of neutrino masses and long proton lifetime several pathways are at our hands, one of them we discussed in the present work, others include obtaining proton decay operator from the neutrino mass origin, and vice versa generating small neutrino masses via the proton decay operators. These and other interesting prospects are to be studied in the future works.
\section{Conclusion}
\label{sec:conclusion}
An enhanced model of scotogenic neutrino mass with naturally long proton lifetime has been described. The existence of dark matter leads to a naturally small neutrino masses and inherently long proton lifetime. Furthermore, the smallness of neutrino masses and proton decay width has a build correlation property. The dark matter abundance has been achieved via dark matter annihilation to SM leptons and quarks with dark matter Yukawa couplings to SM leptons and quarks of the order $\mathcal{O}(0.1)$, with the BSM leptoquark mass value as low as $1200$~GeV. Dark matter direct detection forces the Yukawa coupling of dark matter with the first family quarks to be of the order $\mathcal{O}(0.01)$. The valid range of $2.51\times 10^{-5} < \lambda < 2\times 10^{-3}$ has been obtained for the $\lambda$ coupling in the 2HDM. The scotogenic fermionic dark matter with the mass in the range from $100$~GeV to $10$~TeV successfully satisfies all relevant constraints, including dark matter relic density, direct detection, neutrino mass, proton lifetime, and lepton flavor violation. The enhanced scotogenic neutrino mass paradigm has interconnected the three issues in particle physics: dark matter, smallness of neutrino mass, and stability of the proton.
\acknowledgments
%OP was supported by the National Natural Science Fund of China Grant No.~12350410373. 
The work was supported by the National Natural Science Fund of China Grant No.~12350410373 (O.~P.) and by the Fundamental Research Funds for the Central Universities (T.~N.).
%
%--------------------------------------------------------------------------------------------------------------------------------------------------------------------------------------------------------
%
%\appendix
%
%\section{Appendixes}
%
%\newpage
%\nocite{*}
%
%
%\bibliographystyle{utphys}
\bibliography{references}
\end{document}